\documentclass[superscriptaddress,twocolumn,amsmath,amssymb,showpacs,floatfix]{revtex4-2}

\usepackage[utf8]{inputenc}
\usepackage[usenames,dvipsnames]{xcolor}
\usepackage[english]{babel}
\usepackage[T1]{fontenc}
\usepackage{microtype,ragged2e}

\usepackage{natbib}
\usepackage{appendix,hyphenat,mathtools,dsfont,bm,bbm,amsthm,tikz,pgfplots}
\usepackage[makeroom]{cancel}
\usepackage{lmodern}

\definecolor{mygrey}{gray}{0.35}
\definecolor{myblue}{rgb}{0.2,0.2,0.8}
\definecolor{myzard}{cmyk}{0,0,0.05,0}
\definecolor{mywhite}{rgb}{1,1,1}
\definecolor{myred}{rgb}{0.9,0.1,0.}
\definecolor{dgreen}{rgb}{0.0, 0.5, 0.0}
\usepackage[colorlinks=true,citecolor=myblue,linkcolor=myblue,urlcolor=myblue]{hyperref}
\usepackage[capitalize]{cleveref}
\usepackage{bookmark}

\newcommand{\ket}[1]{\lvert#1\rangle} 

\makeatletter
\def\mbig{\bBigg@{1.2}}
\def\mbigl#1{\mathopen{\bBigg@{1.2}#1}}
\def\mbigr#1{\mathclose{\bBigg@{1.2}#1}}
\renewcommand*\env@matrix[1][*\c@MaxMatrixCols c]{%
   \hskip -\arraycolsep%
   \let\@ifnextchar\new@ifnextchar%
   \array{#1}%
}
\makeatother

\newcommand*\bgcol[2]{{%
   \ifmmode%
      \mathchoice%
         {\colorbox{#1}{$\displaystyle#2$}}%
         {\colorbox{#1}{$\textstyle#2$}}%
         {\colorbox{#1}{$\scriptstyle#2$}}%
         {\colorbox{#1}{$\scriptscriptstyle#2$}}%
   \else%
      \colorbox{#1}{#2}%
   \fi%
}}

\def\d{\mathop{}\!\mathrm d\mathchoice{}{}{\kern-.09em}{\kern-.09em}}



\pgfplotsset{compat=1.14}

\begin{document}

\title{Transfer-tensor description of memory effects in open-system dynamics and multi-time statistics}

\author{Stefano Gherardini}
\email{gherardini@lens.unifi.it}
\affiliation{\mbox{Dipartimento di Fisica e Astronomia, Universit{\`a} degli Studi di Firenze,} via G. Sansone 1, I-50019 Sesto Fiorentino, Italy}
\affiliation{\mbox{LENS, CNR-INO, and QSTAR,} via N. Carrara 1, I-50019 Sesto Fiorentino, Italy}
\affiliation{\mbox{INFN Sezione di Firenze}, via G. Sansone 1, I-50019 Sesto Fiorentino, Italy}

\author{Andrea Smirne}
\email{andrea.smirne@unimi.it}
\affiliation{\mbox{Institute of Theoretical Physics, and IQST, Universit\"at Ulm,} Albert-Einstein-Allee 11, 89069 Ulm, Germany}
\affiliation{Dipartimento di Fisica ``Aldo Pontremoli", Universit{\`a} degli Studi di Milano, via Celoria 16, 20133 Milan,
Italy}
\affiliation{Istituto Nazionale di Fisica Nucleare, Sezione di Milano, via Celoria 16, 20133 Milan,
Italy}

\author{Susana Huelga}
\email{susana.huelga@uni-ulm.de}
\affiliation{\mbox{Institute of Theoretical Physics, and IQST, Universit\"at Ulm,} Albert-Einstein-Allee 11, 89069 Ulm, Germany}

\author{Filippo Caruso}
\email{filippo.caruso@unifi.it}
\affiliation{\mbox{Dipartimento di Fisica e Astronomia, Universit{\`a} degli Studi di Firenze,} via G. Sansone 1, I-50019 Sesto Fiorentino, Italy}
\affiliation{\mbox{LENS, CNR-INO, and QSTAR,} via N. Carrara 1, I-50019 Sesto Fiorentino, Italy}

\begin{abstract}
The non-Markovianity of an arbitrary open quantum system is analyzed in reference to the multi-time statistics given by its monitoring at discrete times. On the one hand, we exploit the hierarchy of inhomogeneous transfer tensors, which provides us with relevant information about the role of correlations between the system and the environment in the dynamics. The connection between the transfer-tensor hierarchy and the CP-divisibility property is then investigated, by showing to what extent quantum Markovianity can be linked to a description of the open-system dynamics by means of the composition of 1-step transfer tensors only. On the other hand, we introduce the set of stochastic transfer tensor transformations associated with local measurements on the open system at different times and conditioned on the measurement outcomes. The use of the transfer-tensor formalism accounts for different kinds of memory effects in the multi-time statistics and allows us to compare them on a similar footing with the memory effects present in non-monitored non-Markovian dynamics, as we illustrate on a spin-boson case study.
\end{abstract}

\maketitle

\section{Introduction}

In a dynamical quantum system the interaction with the external environment typically leads to non-Markovian behaviours, which, broadly speaking, are associated with the presence of memory effects in the system evolution \cite{WolfCMP2008,BreuerPRL2009,LainePRA2010,Rivas2010PRL105,LiuNatPhys2011,ChruscinskiPRL2014,Rivas_Review_2014,CarusoRMP2014,Breuer_Review_2016,MullerSciRep2016,deVegaRMP2017,Cialdi2017,PollockPRL2018,PollockPRA2018,Budini2018,CampbellPRA2018,TarantoPRL2019,DoNJP2019,CampbellNJP2019,Cialdi2019,GherardiniPRR2020,MartinaArXiv2021}.
Although by now many distinct definitions of quantum non-Markovianity have been introduced, two main pathways can be identified \cite{Li2018}.

In one case, the focus is on the evolution of the open-system state at different times, as fixed, e.g., by the dynamical maps or the master equations that characterize the open-system dynamics \cite{Breuer2002,Rivas2012}, while in the other case the focus is on the statistics associated
with a sequence of local measurements (or other active interventions) performed on the open system at subsequent times.
These two approaches to non-Markovianity are indeed inherently different, as the former addresses the predictions related with observables at a single time, while the latter
concerns the multi-time statistics. In addition to such a difference, which already appears at the classical level \cite{Vacchini2011,Smirne2013}, the quantum nature of the system at hand implies a further, fundamental distinction between the single-time and the multi-time
notions of non-Markovianity. Contrary to what happens classically, in quantum systems it is generally not possible, even in principle,
to access the information associated with sequential measurements without disturbing the system that is being measured. Indeed, quantum measurements alter
the current state of the system, as well as its correlations with the environment, which will later
result in a modified evolution \cite{Rivas_Review_2014,Breuer_Review_2016,Smirne2017Coherence,SmirneArxiv2019,Milz2019,Strasberg2019,Milz2020,Diaz2020,MilzQuantum20,proceeding_IQIS}.
As a consequence, the presence of intermediate measurements can bring along specific forms
of memory, which combine in a non-trivial way with the memory strictly due to the system-environment interaction.

In this paper, we investigate the relation between the memory effects appearing in an open quantum system dynamics and those associated with the multi-time statistics due to sequential measurements, by means of the transfer tensor (TT) method\,\cite{Cerrillo2014PRL112,RosenbachNJP2016,Modi_Reconstructing_2018,ChenPRApp2020}.
We will show that the latter, which was introduced to treat efficiently the long-time dynamics
of open quantum systems, also allows one to treat memory effects on the dynamics and the multi-time
statistics on a similar footing. First, we prove to what extent the divisibility property of the dynamical maps,
which is the defining property of quantum non-Markovianity according to\,\cite{Rivas_Review_2014,Rivas2010PRL105}, is linked to the hierarchy of TTs also in the inhomogeneous case, i.e., beyond the time-translational invariant regime explored in the original paper \cite{Cerrillo2014PRL112}. Then, we extend the definition of the TTs to the situation where the open system is measured at subsequent times, via the corresponding conditional stochastic dynamics, and we take into account the multi-time probabilities that define the expectation values
associated with measurements performed at different times. This allows us to identify different forms of memory in the multi-time statistics, related with the interplay between the correlations and environmental-state transformations induced by, respectively, the interaction between the open system and the environment and the sequential measurements. After introducing proper quantifiers related with the $L_2$-norm of the multi-step (i.e., step greater than $1$) TTs, we compare in a case study the relevance of such memory effects for different kinds of measurements, including the case without any intermediate measurement.

The rest of the paper is organized as follows. In Sec.\ref{sec:itt}, we recall the main features of the TTs formalism that are relevant for our analysis,
focusing in particular on the recursive construction of the TTs. In Sec.\ref{sec:ttm}, we prove that if only multi-step TTs are different from zero the open-system dynamics is divisible, while the converse statement does not hold, as we show by means of an explicit example. In Sec.\ref{sec:ctt}, we first generalize the TTs formalism to the case of multi-time measurements, by introducing a family of TTs conditioned on the sequence of measurement outcomes, and then we exploit it to characterize the different kinds of memory effects associated with the multi-time statistics. Finally, the conclusions and outlooks of our analysis are discussed in Sec.\ref{sec:ceo}.

\section{Inhomogeneous transfer tensors hierarchy}\label{sec:itt}

We begin by briefly recalling the TT formalism for the dynamics of an open quantum system \cite{Cerrillo2014PRL112} and the microscopic characterization of the TTs recently derived in \cite{Modi_Reconstructing_2018}, which allows to build up the complete TTs hierarchy and to connect it with the system-environment correlations. Such a construction will be extended to an evolution conditioned on the occurrence of sequential measurements in the following of the paper.

\subsection{General microscopic definition of TTs}

Let us consider a quantum system $\mathbb{S}$ in interaction with an environment $\mathbb{E}$. The dynamics of the composite system $\mathbb{S}+\mathbb{E}$ is governed by a possibly time-dependent Hamiltonian $H(t) = H_0(t) + H_{\rm int}$, where $H_{\rm int}$ takes into account the interaction of the system with the environment, while $H_0(t) = H_{\mathbb{S}}(t) + H_{\mathbb{E}}$ concerns the uncoupled time evolution of $\mathbb{S}$ and $\mathbb{E}$. The interaction Hamiltonian $H_{\rm int}$ is fixed but unknown, and $H_{\mathbb{S}}(t)$ can be time-dependent. Assuming an initial product-state between the open system and the environment and a fixed initial state of $\mathbb{E}$, i.e., $\rho_{t_0} = \rho_{\mathbb{S},t_0} \otimes \rho_{\mathbb{E},t_0}$, the dynamics of $\mathbb{S}$ can be described by means of the formalism of quantum dynamical maps\,\cite{Breuer2002,Rivas2012,CarusoRMP2014}. In fact, one can define a one parameter family of completely positive, trace-preserving (CPTP) dynamical maps $\left\{\Phi(t,t_0)\right\}_{t\geq t_0}$, with $\Phi(t,t_0):\mathcal{S}(\mathcal{H}_{\mathbb{S}})\rightarrow \mathcal{S}(\mathcal{H}_{\mathbb{S}})$, where $\mathcal{S}(\mathcal{H}_{\mathbb{S}})$ denotes the sets of density operators (non-negative operators with unit trace) acting on $\mathcal{H}_{\mathbb{S}}$ and $\Phi(t,t_0)$ is CPTP at any time $t$. Hereafter, the action of the dynamical maps of the system will be taken into account at the discrete time instants (not necessarily equally spaced) $t_{k}$, $k = 1,\ldots,m$, and we will use the notation $\Phi(t_k,t_0)\equiv \Phi_{k}$ and
\begin{equation}\label{dynamical_map}
\rho_{\mathbb{S},k} = \rho_{\mathbb{S}}(t_{k}) = \Phi_{k}[\rho_{\mathbb{S},0}].
\end{equation}
This will make the comparison with the sequential-measurement scenario more transparent and will further allow us to directly apply the TT method \,\cite{Cerrillo2014PRL112}. The latter relies in fact on a family of maps, the TTs $T_{k,j}$, which are defined by the relation
\begin{equation}\label{tt1}
T_{k,0} = \Phi_{k} - \sum_{j = 1}^{k - 1}T_{k,j}\Phi_{j},
\end{equation}
equivalently expressed by
\begin{equation}\label{tt2}
\Phi_{k} = \sum_{j = 0}^{k - 1} T_{k,j}\Phi_{j}.
\end{equation}
In other terms, the state at the time instant $t_k$ is obtained by propagating the states at $t_j$ according to the equation $\rho_{\mathbb{S},k} = \sum_{j=0}^{k-1} T_{k,j}\rho_{\mathbb{S},j}$. We stress that, at variance with the original formulation, we are not assuming time-invariance; thus our analysis will also apply to time-dependent Hamiltonians and non-stationary initial environmental states\,\cite{Cerrillo2014PRL112}.

A useful expression for generic, possibly inhomogeneous TTs was recently derived in\,\cite{Modi_Reconstructing_2018}. Given the global unitary dynamics of the composite system $\mathbb{S}+\mathbb{E}$ from $t_{s}$ to $t_{k}$, i.e., $\mathcal{U}_{t_s:t_k}[\rho] \equiv U_{t_s:t_k}\rho\,U_{t_s:t_k}^{\dagger}$ with $U_{t_s:t_k} \equiv \mathcal{T}\exp(-i\int_{t_s}^{t_k}H(\tau)d\tau)$ ($\mathcal{T}$ is the time ordering operator), let us introduce the CPTP maps $\Gamma_{k|k-n}:\mathcal{S}(\mathcal{H}_{\mathbb{S}})\rightarrow \mathcal{S}(\mathcal{H}_{\mathbb{S}})$ defined as
\begin{equation}
\Gamma_{k|k-n}[\sigma_{\mathbb{S}}] \equiv {\rm Tr}_{\mathbb{E}}\left[\mathcal{U}_{t_{k-n}:t_k}[\sigma_{\mathbb{S}}\otimes\sigma_{\mathbb{E},k-n}]\right],\label{Gamma_main_text}
\end{equation}
where ${\rm Tr}_{\mathbb{S}/\mathbb{E}}[\cdot]$ denotes the partial trace w.r.t.\,the Hilbert space of $\mathbb{S}$ or $\mathbb{E}$, $\sigma_{\mathbb{S}}$ denotes a density operator on $\mathcal{H}_{\mathbb{S}}$ and $\sigma_{\mathbb{E},k-n}$ is a time-dependent density operator on $\mathcal{H}_{\mathbb{E}}$. Here, we will focus on the case where $\sigma_{\mathbb{E},k-n}$ is the state of the environment at $t_{k-n}$. It can then be proven that the $n-$step TT $T_{k,k-n}$ is given by the recursive relation\,\cite{Modi_Reconstructing_2018}
\begin{equation}\label{hierarchy}
T_{k,k-n} = \Gamma_{k|k-n} - \sum_{j=1}^{n-1}T_{k,k-j}\Gamma_{k-j|k-n}
\end{equation}
that allows one to reconstruct the whole hierarchy of TTs from the CPTP maps $\Gamma_{k|k-n}$, as shown explicitly in\,\cite{Modi_Reconstructing_2018}. In the next paragraph, we repeat the construction for the lowest orders of the hierarchy, both for the sake of illustration and since it will be useful in the following when we will move to the stochastic case.

\subsection{Reconstruction of the TTs hierarchy}

First, let us combine together the definition of the dynamical maps of the system at the discrete time instants $t_k$, $k=1,\ldots,n$, i.e., Eq.\,(\ref{dynamical_map}), and the recursive general expression of the TT transformation given by Eq.(\ref{tt2}). In this way, the latter can be written via the following recursive relation:
\begin{equation}\label{TT_recursion}
\Phi_{k} = \underline{T}_{k}\underline{\Xi}_{k - 1}[\rho_{\mathbb{S},0}],
\end{equation}
where
\begin{equation*}
\underline{T}_{k} \equiv \left( T_{k,0}, T_{k,1}, \ldots, T_{k,k-1} \right)
\end{equation*}
and
\begin{equation*}
\underline{\Xi}_{k - 1}[\rho_{\mathbb{S},0}] \equiv
\begin{pmatrix}
\mbox{Id}_{\mathbb{S}}[\rho_{\mathbb{S},0}] \\
\Phi_1 = \underline{T}_{1}\underline{\Xi}_{0}[\rho_{\mathbb{S},0}] \\
\vdots \\
\Phi_{k-1} = \underline{T}_{k-1}\underline{\Xi}_{k-2}[\rho_{\mathbb{S},0}]
\end{pmatrix},
\end{equation*}
which are valid for $n \geq 1$ ($\underline{T}_{0} = \underline{\Xi}_{- 1} = \mbox{Id}_{\mathbb{S}}[\rho_{\mathbb{S},0}]$, where $\mbox{Id}_{\mathbb{S}}$ denotes the identity map on
the set of operators on $\mathbb{S}$). For the sake of clarity, we show the first terms of the recursive expansion of Eq.\,(\ref{TT_recursion}):
\begin{equation}\label{eq:cases}
\begin{cases}
\Phi_{0} = \mbox{Id}_{\mathbb{S}}[\rho_{\mathbb{S},0}] \\
\Phi_{1} = T_{1,0}\Phi_{0} = T_{1,0}[\rho_{\mathbb{S},0}] \\
\Phi_{2} = T_{2,0}\Phi_{0} + T_{2,1}\Phi_{1} = \left(T_{2,0}+T_{2,1}T_{1,0}\right)[\rho_{\mathbb{S},0}] \\
\Phi_{3} = T_{3,0}\Phi_{0} + T_{3,1}\Phi_{1} + T_{3,2}\Phi_{2} \\
\,\,\,\,\,\,\,\,\, = \left(T_{3,0} + T_{3,1}T_{1,0} + T_{3,2}T_{2,0} + T_{3,2}T_{2,1}T_{1,0}\right)[\rho_{\mathbb{S},0}]. \\
\end{cases}
\end{equation}
By combining Eqs.\,(\ref{dynamical_map}) and (\ref{TT_recursion}), we immediately find that
\begin{equation}\label{T_10}
T_{1,0}[\rho_{\mathbb{S},0}] = \Phi_{1}[\rho_{\mathbb{S},0}],
\end{equation}
so that $\rho_{\mathbb{S},1} = T_{1,0}[\rho_{\mathbb{S},0}]$. Then, by continuing the recursion, one has
\begin{equation*}
\rho_{\mathbb{S},2} = T_{2,0}[\rho_{\mathbb{S},0}] + T_{2,1}[T_{1,0}[\rho_{\mathbb{S},0}]].
\end{equation*}
Here, it is worth observing that $T_{2,1}[T_{1,0}[\rho_{\mathbb{S},0}]]$ returns the component of $\rho_{\mathbb{S},2}$ conditioned on finding the system in the states $\rho_{\mathbb{S},0}$ and $T_{1,0}[\rho_{\mathbb{S},0}]$ at the time instants $t_0$ and $t_1$ and no system-environment correlations were present.
In fact, given the maps $\Gamma_{k|k-n}$ defined in Eq.\,(\ref{Gamma_main_text}), the 1-step TTs $T_{k,k-1}$ satisfy
\begin{equation}\label{eq:onest}
T_{k,k-1} = \Gamma_{k|k-1}
\end{equation}
as a direct consequence of Eq.\,(\ref{hierarchy}) (for the trivial case $n=1$). In other terms, we have
\begin{equation}\label{eq:rho2}
T_{2,0}[\rho_{\mathbb{S},0}] = \rho_{\mathbb{S},2} - \Gamma_{2|1}[\rho_{\mathbb{S},1}],
\end{equation}
i.e.,
\begin{equation}\label{T20}
T_{2,0} =  \Phi_{2} - \Gamma_{2|1}\Phi_1.
\end{equation}
For the sake of brevity, from here on we will remove the subscript $\mathbb{S}$ from the notation; it will used (together with $\mathbb{E}$) only if necessary.\\
Moving on to the level $k = 3$ of the hierarchy, by applying Eq.\,(\ref{eq:cases}) one has that the reduced density operator of $\mathbb{S}$ at $t_3$ is equal to
\begin{eqnarray*}
\rho_{3} &=& T_{3,0}[\rho_{0}] + T_{3,1}[T_{1,0}[\rho_{0}]] + T_{3,2}[T_{2,0}[\rho_{0}]]\nonumber\\
&+& T_{3,2}[T_{2,1}[T_{1,0}[\rho_{0}]]].
\end{eqnarray*}
Using the identity in Eq.\,(\ref{eq:onest}) for $T_{3,2}$ and $T_{2,1}$, as well as Eq.\,(\ref{eq:rho2}), we have
\begin{equation}\label{T_32T_20}
\begin{cases}
T_{3,2}[T_{2,0}[\rho_{0}]] = \Gamma_{3|2}[\rho_{2} - \Gamma_{2|1}[\rho_{1}]] \\
T_{3,2}[T_{2,1}[T_{1,0}[\rho_{0}]]] = \Gamma_{3|2}\Gamma_{2|1}[\rho_{1}].
\end{cases}
\end{equation}
Moreover, $T_{3,1}$ can be expressed by exploiting Eq.\,(\ref{hierarchy}) (for $n=2$) that along with Eq.\,(\ref{eq:onest}) gives
\begin{equation}\label{T31}
T_{3,1}\equiv \Gamma_{3|1}- \Gamma_{3|2}\Gamma_{2|1}.
\end{equation}
Thus, $T_{3,1}$ can be written as a function of a term involving only one-step $\Gamma$s, i.e., $\Gamma_{3|2}\Gamma_{2|1}$, and of the $2-$step $\Gamma_{3|1}$. By combining together all terms of Eqs.\,(\ref{T_32T_20}) and (\ref{T31}), one has
\begin{equation*}
\rho_{3} = T_{3,0}[\rho_{0}] + \Gamma_{3|1}[\rho_{1}] + \Gamma_{3|2}[\rho_{2}] - \Gamma_{3|2}\Gamma_{2|1}[\rho_{1}],
\end{equation*}
i.e.,
\begin{equation}\label{T_30}
T_{3,0} = \Phi_3 - \Gamma_{3|2}\Phi_2 + (\Gamma_{3|2}\Gamma_{2|1} - \Gamma_{3|1})\Phi_1.
\end{equation}

As mentioned, the procedure can be generalized to any $k$ via Eq.\,(\ref{hierarchy}), so that the whole hierarchy of TTs can be expressed in terms of the maps $\Gamma_{k|k-n}$\,\cite{Modi_Reconstructing_2018} (indeed, $\Phi_k = \Gamma_{k|0}$). Importantly, this construction reveals the influence of the correlations between the system and the environment into the dynamics of the open quantum system. In fact, the CPTP maps (\ref{Gamma_main_text}) generate a hierarchy of $1-$, $2-$,...,$n-$step TTs conditioned to the fact that the system passes through product-states at the different steps taken into account. As we will show in the next section, this is naturally linked to the presence of memory effects in the open-system dynamics.

\section{Transfer tensors and Markovianity of the dynamics}\label{sec:ttm}

Here, we derive a definite connection between a property of the TTs -- the possibility to build up the whole hierarchy only in terms of $1-$step TTs -- and the non-Markovianity of the dynamics. In particular, we rely on the definition introduced in\,\cite{Rivas2010PRL105}, which identifies Markovian dynamics as those described by a family of CP-divisible dynamical maps. Adapting the original definition in\,\cite{Rivas2010PRL105} to the case of a discrete set of times, then the following definition can be stated: \\
The (discrete) dynamics $\left\{\Phi_k\right\}_{k=1,\ldots,m}$ is Markovian when for any $k\geq j\geq 0$ there is a CPTP map $\mathcal{E}_{k,j}$ such that $\Phi_k= \mathcal{E}_{k,j}\Phi_j$.

Hence, if we assume that $T_{k,k-n}=0$ for $n\geq 2$, then by applying recursively Eq.\,(\ref{tt2}) we have
\begin{equation}
\Phi_k = T_{k,k-1}T_{k-1,k-2}\cdots T_{j+1,j}\Phi_j
\end{equation}
for any $k\geq j\geq0$. But then Eq.\,(\ref{eq:onest}) implies
\begin{equation}\label{eq:aux}
\Phi_k = \Gamma_{k|k-1}\Gamma_{k-1|k-2}...\Gamma_{j+1|j}\Phi_j.
\end{equation}
Any conditional map $\Gamma_{k|j}$ is CPTP by construction, see Eq.(\ref{Gamma_main_text}), so that we can conclude that any map $\Phi_k$ can be decomposed
as $\Phi_{k}=\mathcal{E}_{k, j} \Phi_j$ with $\mathcal{E}_{k,j}$ CPTP map: \\
\emph{If only one-step TTs are different from zero, the resulting (discrete-time) evolution is CP-divisible}. \\
We further note that if we restrict to the case of equally-spaced time instants, $t_k = k \Delta$, and translational invariant TTs (i.e., $T_{k,k-n}= T_{n,0}$), then $T_{k,k-n}=0$ for $n\geq 2$ implies that the open-system dynamics is not only Markovian, but also a semigroup, i.e., $\Phi_k = (\Phi_1)^k$, thus recovering what shown in\,\cite{Cerrillo2014PRL112}.

The previous result, besides linking a property of the hierarchy of TTs to the Markovianity of the corresponding dynamics, provides us with an explicit illustration
of one of the physical meanings of such property. In fact, from Eq.(\ref{eq:aux}) we see that when one-step TTs are the only non-zero TTs, the dynamical maps can be obtained by using the maps $\Gamma$ only. System-environment correlations, despite being present as a consequence of the interaction term within the unitary operators, will not affect the reduced dynamics of the open system, so that at any time $t_{k-n}$ the actual global state can be effectively replaced by a product-state $\rho_{\mathbb{S},k-n}\otimes \sigma_{\mathbb{E},k-n}$, see Eq.(\ref{Gamma_main_text}). Experimentally, Eq.\,(\ref{eq:aux}) can be validated by independently reconstructing the maps $\{\Phi_k\}$ and the 1-step $\{\Gamma_{k|k-1}\}$, obtained by breaking the system-environment correlations at the time instants $t_k$. In particular, the former can be reconstructed by means of quantum tomography processes, while the latter, ideally, by preparing a fresh copy of the system in the state $\sigma_{\mathbb{S},k}$\,\cite{PollockPRL2018,PollockPRA2018}, which is known by virtue of the previously reconstructed map $\Phi_k$. Alternatively, and more realistically in general scenarios, the map $\Gamma_{k|k-1}$ can be reconstructed by exploiting its linearity, a complete set of projective measurements
$\left\{\mathcal{P}_i = |u_i\rangle\!\langle u_i|\right\}_{i=1,\dots,d_{\mathbb{S}}}$ on $\mathbb{S}$ (with $d_{\mathbb{S}}$ the dimension of $\mathcal{H}_{\mathbb{S}}$), where $\left\{\ket{u_i}\right\}_{i=1,\dots,d_{\mathbb{S}}}$  is an orthonormal basis on $\mathcal{H}_{\mathbb{S}}$, and a set of local operations $\left\{\mathcal{R}_{i j}\right\}_{j=1,\ldots,d_{\mathbb{S}}^2}$ such that $\left\{\mathcal{R}_{i j}\left[|u_i\rangle\!\langle u_i|\right]\right\}_{j=1,\ldots,d_{\mathbb{S}}^2}$ is a linearly independent set of system $\mathbb{S}$ states. In fact, let $\Gamma^{(i)}_{k|k-1}$ be the map defined as
\begin{equation}
\Gamma^{(i)}_{k|k-1}[\sigma_{\mathbb{S}}] \equiv {\rm Tr}_{\mathbb{E}}\left[\mathcal{U}_{t_{k-1}:t_k}[\sigma_{\mathbb{S}}\otimes\sigma^{(i)}_{\mathbb{E},k-1}]\right],
\end{equation}
where $\sigma^{(i)}_{\mathbb{E},k-1} = {\rm Tr}_{\mathbb{S}}\left[(\mathcal{P}_{i} \otimes \mbox{Id}_{\mathbb{E}})[\rho_{\mathbb{S}\mathbb{E}, k-1}]\right]$ with $\mbox{Id}_{\mathbb{E}}$ denoting the identity map on the set of operators on $\mathbb{E}$. Such a map can be tomographically reconstructed by standard techniques: one has to perform the projective measurement $\mathcal{P}_i$ at the $(k-1)$-th time step, followed by one of the local operations $\mathcal{R}_{i j}$ that will not alter the
subnormalized density operator $\sigma^{(i)}_{\mathbb{E},k-1}$. In this way, we can reconstruct the quantum state $\Gamma^{(i)}_{k|k-1}\left[\mathcal{R}_{ij}\left[|u_i\rangle\!\langle u_i|\right]\right]$ for each $j=1,\ldots,d_{\mathbb{S}}^2$, and then, the map $\Gamma_{k|k-1}$ is obtained by repeating the procedure for each $\mathcal{P}_i$, with $i=1,\dots,d_{\mathbb{S}}$. Indeed, by linearity $\sigma_{\mathbb{E},k-1}=\sum^{d_{\mathbb{S}}}_{i=1} \sigma^{(i)}_{\mathbb{E},k-1}$
and then $\Gamma_{k|k-1}=\sum^{d_{\mathbb{S}}}_{i=1}\Gamma^{(i)}_{k|k-1}$. For all purposes, the reconstruction procedure argued above can be achieved experimentally by means of a quantum process tomography, which however suffers of the drawback to require lots of classical and quantum resources scaling quadratically with the dimension of $\mathbb{S}$ (see e.g.\,Ref.\,\cite{NielsenBook2000}, chapter 10). Otherwise, even a randomized benchmarking protocol \cite{MagesanPRL2011} could be taken into account as possible solution, though further investigations are needed.

The fact that only one-step TTs are non-zero is in general a \textit{stronger} requirement than the CP-divisibility of the dynamics, as follows from the analysis of \cite{Li2018} (see also \cite{foot1}) and the explicit example provided below. Indeed, having only one-step TTs different from zero is only a \emph{sufficient criterion} (and thus not necessary) for CP-divisible discrete-time evolutions. In this regard, let us consider the simplest case, namely $m=2$, whereby we recall that $m$ is the last (and greater) value of the index $k$ in the sequence of time instants $t_k$. Hence, the only non-trivial requirement for CP-divisibility is the existence of a CPTP map $\mathcal{E}_{2,1}$ obeying the relation
\begin{equation*}
\mathcal{E}_{2,0} = \Phi_{2} = \mathcal{E}_{2,1}\Phi_{1} = \mathcal{E}_{2,1}T_{1,0}
\end{equation*}
since by definition $T_{1,0} \equiv \Phi_{1}$. Now, according to the definition of the TT transformation, $\Phi_{2}$ is also equal to
\begin{equation*}
\Phi_{2} = T_{2,0} + T_{2,1}T_{1,0}.
\end{equation*}
Therefore, assuming that the inverse of $T_{1,0}$ (i.e., $(T_{1,0})^{-1}$) does exist, one has
\begin{equation}
\mathcal{E}_{2,1} = T_{2,0}\left(T_{1,0}\right)^{-1} + T_{2,1}.
\end{equation}
What we will show is that $\mathcal{E}_{2,1}$ can be CPTP also if $ T_{2,0}\neq 0$ and even if $\left(T_{1,0}\right)^{-1}$ is not be generally CPTP. In fact, consider the reduced dynamics of a two-level system, fixed by the unitary dynamics given by the global Hamiltonian $H = \sigma_z+H_\mathbb{E}+\sigma_z \otimes B$, where the free environmental Hamiltonian $H_\mathbb{E}$ and the interaction term commute: $[H_\mathbb{E}, B] = 0$. Now, also assume that the global state at the initial time $t_0=0$ is $\rho(0)=\rho_\mathbb{S}(0) \otimes \rho_\mathbb{E}(0)$, where the environmental state is stationary with respect to the free dynamics such that $[H_\mathbb{E}, \rho_\mathbb{E}] = 0$. As a consequence,
on the one hand, the reduced dynamics of the open system is a pure dephasing, namely $\rho_{11}(t) = \rho_{11}(0)$ and $\rho_{10}(t)=k(t) \rho_{10}(0)$, where $\rho_{ij}(t) = \langle i| \rho_\mathbb{S}(t)|j\rangle$ ($\left\{\ket{1}, \ket{0}\right\}$ are the eigenvectors of $\sigma_z$) and $k(t) = e^{i \omega t}\mbox{Tr}_{\mathbb{E}}[e^{2 i B t} \rho_\mathbb{E}]$. On the other hand, the state of the environment will not evolve in time, also when interacting with the system, i.e., $\rho_\mathbb{E}(t) = \rho_\mathbb{E}(0)$. For any two times $t_2\geq t_1\geq0$, we will then have
\begin{equation*}
T_{2,1}=\Gamma_{2|1}=\Phi_1,
\end{equation*}
where we used Eq.(\ref{eq:onest}) and Eq.(\ref{Gamma_main_text}), respectively. As a result (see also Eq.(\ref{T20})),
\begin{equation*}
T_{2,0}=\Phi_2-\Phi^2_1.
\end{equation*}
Specifically, using the matrix representation in the basis $\left\{\ket{1}, \ket{0}\right\}$, for a generic
state $\rho_{\mathbb{S}}$ we have
$$
\mathcal{E}_{2,1}[\rho_{\mathbb{S}}] = \begin{pmatrix}
\rho_{11}& \frac{k(t_2)}{k(t_1)}\rho_{10}\\
\frac{k^*(t_2)}{k^*(t_1)}\rho_{01}& \rho_{00}
\end{pmatrix},
$$
which is positive definite for any $\rho_{\mathbb{S}}$ if and only if $|k(t_2)| \leq |k(t_1)|$. Note that this implies that $\mathcal{E}_{2,1}$ is also CPTP under the same condition, since for pure dephasing positivity and complete positivity coincide. On the other hand,
$$
T_{2,0}[\rho_{\mathbb{S}}]
=\begin{pmatrix}
0 &(k(t_2) -k(t_1)^2)\rho_{10}\\
(k^*(t_2)-k^*(t_1)^2)\rho_{01}&0
\end{pmatrix}
$$
that is equal to 0 for any $\rho_{\mathbb{S}}$ if and only if $k(t_2)=k(t_1)^2$. The latter condition implies but is not implied by the former (indeed, $|k(t)|\leq 1$), which proves our claim: $\mathcal{E}_{2,1}$ can be CPTP, and hence the dynamics CP-divisible,
even though the $2$-step transfer tensor $T_{2,0}$ is different from zero.

\section{Transfer tensors and multi-time statistics}\label{sec:ctt}

In this paragraph, we move on to the second part of our investigation, where we consider an open system that, besides interacting with the environment, is measured at some discrete instants of time. We will first introduce a proper counterpart of the transfer tensors in such dynamical regime, and will then show how it can be used to account for the memory present in the multi-time statistics. Before that, let us specify the framework we refer to.

Assume that the open quantum system $\mathbb{S}$ is monitored at the same $m$ instants of time where the discrete dynamics has been evaluated so far. In particular, we take a sequence of quantum measurements locally performed on $\mathbb{S}$ according to the observables $\mathcal{O}_k \equiv F_{\theta_k} \otimes \mathbbm{1}_{\mathbb{E}}$, where $\mathbbm{1}_{\mathbb{E}}$ is the identity operator on $\mathcal{H}_{\mathbb{E}}$, $\{\theta_k\}$ is the set of the possible measurement outcomes, and $\{F_{\theta_k}\}$ denotes the set of positive semi-definite operators on $\mathcal{H}_{\mathbb{S}}$ satisfying the relation $\sum_{\theta_k}F_{\theta_k} = \mathbbm{1}_{\mathbb{S}}$ $\forall k$. The probability that the outcome $\theta_k$ associated with the measurement operator $F_{\theta_k}$ occurs is equal to ${\rm Tr}[\rho_{\mathbb{S},k}F_{\theta_k}]$, while the post-measurement state of $\mathbb{S}$ equals to $\widehat{\rho}_{\mathbb{S},k} \equiv \mathcal{M}_{\theta_{k}}[\rho_{\mathbb{S},k}] /{\rm Tr}[\mathcal{M}_{\theta_{k}}[\rho_{\mathbb{S},k}]]$. It is worth observing that $\mathcal{M}_{\theta_{k}}$ is a CP and trace non-increasing map, while $\sum_{\theta_{k}}\mathcal{M}_{\theta_{k}}$ is a CPTP map; indeed, this corresponds to a quantum instrument, i.e., the most general description of a measurement-induced transformation according to the rules of quantum mechanics \cite{Heinosaari2012}. Specifically, due to CP, any map $\mathcal{M}_{\theta_{k}}$ can be written as $\mathcal{M}_{\theta_{k}}[\rho_{\mathbb{S},k}] \equiv \sum_{i_k}M_{\theta_{k},i_k}\rho_{\mathbb{S},k}M^{\dagger}_{\theta_{k},i_k}$ where the operators $M_{\theta_k,i_k}$ fulfill the identity $F_{\theta_k} = \sum_{i_k}M^{\dagger}_{\theta_k,i_k}M_{\theta_k,i_k}$. Moreover, also note that $\rho_{\mathbb{S},k}$ ($\widehat{\rho}_{\mathbb{S},k}$) denotes the state of the system before (after) the measurement at the
time step $k$.

Importantly, a measurement of the open system affects not only its current state, but also its future evolution as a consequence of the change in the correlations between the system and the environment due to the measurement itself\,\cite{LuchnikovPRL2020}. The very notion of dynamics of the open system becomes more subtle,
since it cannot be clearly separated from the results of the sequential measurements. It is thus useful to introduce the notion of \textit{conditional dynamics}, whereby the system admits a different dynamics for each sequence of measurement outcomes. Explicitly, the conditional dynamics of the system (for a given global evolution $\mathcal{U}_{t_s:t_k}$ and initial environmental state $\rho_{\mathbb{E},0}$) is fixed by the CP map
\begin{eqnarray}\label{eq:phit}
\widetilde{\Phi}^{\underline{\theta},\underline{t}}_k[\rho_{\mathbb{S},0}] &\equiv& {\rm Tr}_{\mathbb{E}}
\left[\mathcal{U}_{t_{k-1}:t_k}(\mathcal{M}_{\theta_{k-1}}\otimes\mbox{Id}_{\mathbb{E}})\cdots\right.\nonumber \\
&\cdots&\left.(\mathcal{M}_{\theta_1}\otimes\mbox{Id}_{\mathbb{E}})\,\mathcal{U}_{t_0:t_1}[\rho_{\mathbb{S},0} \otimes \rho_{\mathbb{E},0}]\right],
\end{eqnarray}
depending on the time instants $\underline{t} \equiv (t_1, \ldots, t_k)$ as well as on the measurement outcomes $\underline{\theta} \equiv (\theta_1,\ldots,\theta_{k-1})$. In other terms, $\widetilde{\Phi}^{\underline{\theta},\underline{t}}_k$ has to be understood as a \textit{stochastic} map, so that we could effectively define different trajectories in the set of CP maps, each of them associated with a different sequence of measurement outcomes. In general $\widetilde{\Phi}^{\underline{\theta},\underline{t}}_k$ is not trace preserving.

The joint probability distributions to get the measurement outcomes $\theta_1,\theta_2,\ldots,\theta_k$ at the time instants $t_1,t_2,\ldots,t_k$ is directly linked to the stochastic map $\widetilde{\Phi}^{\underline{\theta},\underline{t}}_k$ by the following relation:
\begin{equation}\label{joint_prob}
q_k \equiv {\rm Prob}(\theta_{k},t_{k};\ldots;\theta_{1},t_{1}) = {\rm Tr}[\mathcal{M}_{\theta_k}\widetilde{\Phi}^{\underline{\theta},\underline{t}}_{k}[\rho_{\mathbb{S},0}]].
\end{equation}
These quantities define the multi-time statistics associated to sequential measurements at different times and, as recalled in the introduction, suitable notions of Markovianity can be attributed to them\,\cite{Smirne2017Coherence,Li2018,Budini2018,PollockPRL2018,PollockPRA2018,Milz2019,Strasberg2019,Milz2020,MilzQuantum20}. Importantly, from the multi-time probabilities defined in Eq.\,(\ref{joint_prob}), one can reconstruct the expectation value of any sequence of observables at different times, which are of direct interest in experimental applications. In fact, let $\mathcal{O}_1, \ldots \mathcal{O}_k$ be the observables with possible outcomes $\left\{\theta_1\right\}, \ldots \left\{\theta_k\right\}$; hence,
\begin{eqnarray}
\langle \mathcal{O}_1(t_1) \ldots \mathcal{O}_k(t_k)\rangle &=& \sum_{\theta_1,\ldots,\theta_k}
\theta_1\ldots\theta_k{\rm Prob}(\theta_{k},t_{k};\ldots;\theta_{1},t_{1}) \nonumber\\
&=& \sum_{\theta_1,\ldots,\theta_k} q_k\,
\theta_1\ldots\theta_k
\end{eqnarray}
provides us with the mentioned expectation value.

\subsection{Conditional transfer tensors}

As anticipated, in order to characterize the conditional dynamics originated by different sequences of outcomes, we can introduce a stochastic version of the TTs. Let us denote them as $\widetilde{T}^{\underline{\theta},\underline{t}}_{k,j}$, with $k,j = 1,\ldots,m$, explicitly pointing out their dependence on the instants and outcomes of the repeated measurements.

The basic idea is to express the definition of the conditional dynamical map in a recursive fashion. As first step, note that the global state after the first measurement is proportional (via the normalization factor) to
\begin{equation}
\widehat{\rho}_{1} \sim \left(\mathcal{M}_{\theta_1}\otimes \mbox{Id}_E\right)\mathcal{U}_{t_0:t_1}\left[\rho_{0}\right] \equiv \mathcal{V}_{t_0:t_1}[\rho_{0}]
\end{equation}
that also provides us the formal definition of the map $\mathcal{V}[\rho]$. Accordingly, the stochastic quantum map of $\mathbb{S}$ at time $t_1$, i.e., $\widetilde{\Phi}_1$, can be implicitly linked to $\mathcal{V}_{t_0:t_1}[\rho_{0}]$ through the relation
\begin{equation}
\mathcal{M}_{\theta_{1}}\widetilde{\Phi}_{1}[\rho_{\mathbb{S},0}] \equiv {\rm Tr}_{\mathbb{E}}\left[\mathcal{V}_{t_0:t_1}[\rho_{0}]\right].
\end{equation}
Analogously, the $k-$th dynamical map of $\mathbb{S}$ after a sequence of quantum measurements is defined by
\begin{equation}
\widehat{\rho}_{k} \sim  \left(\mathcal{M}_{\theta_k}\otimes \mbox{Id}_E\right)\mathcal{U}_{t_{k-1}:t_{k}}\left[\widehat{\rho}_{k-1}\right]
\equiv \mathcal{V}_{t_{k-1}:t_{k}}[\widehat{\rho}_{k-1}]
\end{equation}
with
\begin{equation}\label{eq:ytr}
\mathcal{M}_{\theta_{k}}\widetilde{\Phi}_{k}[\rho_{\mathbb{S},0}] = {\rm Tr}_{\mathbb{E}}\left[\mathcal{V}_{t_{k-1}:t_{k}}[\widehat{\rho}_{k-1}]\right]
\end{equation}
and $\widehat{\rho}_{0} \equiv \rho_{0}$. For the sake of clarity, we recall that, by definition, the CP maps $\widetilde{\Phi}_{k}$ are obtained by tracing the state of the composite system $\mathbb{S}+\mathbb{E}$ (w.r.t.\,the environment $\mathbb{E}$) just after applying the evolution map $\mathcal{U}_{t_{k-1}:t_k}$. As done with the definition of transfer tensors linking together the quantum maps $\Phi_k$ of the system at time instants $t_k$, $k=1,\ldots,m$, we can thus introduce the stochastic transformations $\widetilde{T}^{\underline{\theta},\underline{t}}_{k,j}$ that relate the conditional dynamics of $\mathbb{S}$ after each measurement of the sequence. In particular, Eq.\,(\ref{eq:ytr}) has the same structure as the not monitored dynamics, with $\Phi_k$ replaced by $\mathcal{M}_{\theta_{k}}\widetilde{\Phi}_{k}$. Hence, we define the stochastic TTs via
\begin{equation}\label{transfer_tensor_transformation_app}
\widetilde{\Phi}^{\underline{\theta},\underline{t}}_k = \sum_{j = 0}^{k - 1}\widetilde{T}^{\underline{\theta},\underline{t}}_{k,j}\mathcal{M}_{\theta_{j}}\widetilde{\Phi}^{\underline{\theta},\underline{t}}_j\,.
\end{equation}
Note that if there are no measurements, $\mathcal{M}_{\theta_{k}}=\mbox{Id}$ and $\widetilde{\Phi}^{\underline{\theta},\underline{t}}_k=\Phi_k$ (see Eq.\,(\ref{eq:phit})) with the result that the TTs $\widetilde{T}^{\underline{\theta},\underline{t}}_{k,j}$ are no longer conditional objects and can be identified with the $T_{k,j}$ since Eq.\,(\ref{transfer_tensor_transformation_app}) reduces to Eq.\,(\ref{tt2}). From now on, we will drop the label $(\cdot)^{\underline{\theta},\underline{t}}$, which is implied in all the expressions with a tilde.

As before, by expanding Eq.\,(\ref{transfer_tensor_transformation_app}) we can write the expression for the $k-$th dynamical map of $\mathbb{S}$ as a function of $\rho_{\mathbb{S},0}$, i.e.,
\begin{equation}\label{TTM_measure_2}
\widetilde{\Phi}_k = \underline{\widetilde{T}}_{k}\underline{\mathcal{M}}_{\theta_{k - 1}}\underline{\Upsilon}_{k - 1}[\rho_{\mathbb{S},0}],
\end{equation}
where
\begin{eqnarray}
\underline{\widetilde{T}}_{k} &\equiv& \left( \widetilde{T}_{k,0}, \widetilde{T}_{k,1}, \ldots, \widetilde{T}_{k,k-1} \right) \nonumber\\
\underline{\mathcal{M}}_{\theta_{k - 1}} &\equiv& {\rm diag}\left(\mathcal{M}_{\theta_{0}},\mathcal{M}_{\theta_{1}},\ldots,\mathcal{M}_{\theta_{k-1}}\right)
\end{eqnarray}
and
\begin{equation}
\underline{\Upsilon}_{k - 1}[\rho_{\mathbb{S},0}] \equiv \begin{pmatrix}
\mbox{Id}_{\mathbb{S}}[\rho_{\mathbb{S},0}] \\
\widetilde{\Phi}_1 = \underline{\widetilde{T}}_{1}\underline{\Upsilon}_{0}[\rho_{\mathbb{S},0}] \\
\widetilde{\Phi}_2 = \underline{\widetilde{T}}_{2}\underline{\mathcal{M}}_{\theta_{1}}\underline{\Upsilon}_{1}[\rho_{\mathbb{S},0}] \\
\vdots \\
\widetilde{\Phi}_{k-1} = \underline{\widetilde{T}}_{k-1}\underline{\mathcal{M}}_{\theta_{k-2}}\underline{\Upsilon}_{k-2}[\rho_{\mathbb{S},0}]
\end{pmatrix}
\end{equation}
with $\underline{\Upsilon}_{- 1}$ and $\mathcal{M}_{\theta_{0}}$ equal to the identity map
$\mbox{Id}_{\mathbb{S}}$ (the first measurement of the sequence, indeed, is not performed at time $t_0$, but at $t_1$) and $\underline{\Upsilon}_{0} = \underline{\Xi}_{0} = \Phi_0$.

For the sake of clarity, we show also in this case the first terms of the recursive expansion of Eq.\,(\ref{TTM_measure_2}):

\begin{small}
\begin{equation*}
\begin{cases}
\widetilde{\Phi}_{0} = \Phi_{0} = \mbox{Id}_{\mathbb{S}}[\rho_{\mathbb{S},0}] \\
\widetilde{\Phi}_{1} = \widetilde{T}_{1,0}\widetilde{\Phi}_{0} = \widetilde{T}_{1,0}[\rho_{\mathbb{S},0}] \\
\widetilde{\Phi}_{2} = \widetilde{T}_{2,0}\widetilde{\Phi}_{0} + \widetilde{T}_{2,1}\mathcal{M}_{\theta_{1}}\widetilde{\Phi}_{1} = \left(\widetilde{T}_{2,0}+\widetilde{T}_{2,1}\mathcal{M}_{\theta_{1}}\widetilde{T}_{1,0}\right)[\rho_{\mathbb{S},0}] \\
\widetilde{\Phi}_{3} = \widetilde{T}_{3,0}\widetilde{\Phi}_{0} + \widetilde{T}_{3,1}\mathcal{M}_{\theta_{1}}\widetilde{\Phi}_{1} + \widetilde{T}_{3,2}\mathcal{M}_{\theta_{2}}\widetilde{\Phi}_{2} \\
\,\,\,\,\,\,\,\,\, = \left(\widetilde{T}_{3,0} + \widetilde{T}_{3,1}\mathcal{M}_{\theta_{1}}\widetilde{T}_{1,0}+\widetilde{T}_{3,2}\mathcal{M}_{\theta_{2}}\widetilde{T}_{2,0}\right.  \\
\,\,\,\,\,\,\,\, \left. +\,\widetilde{T}_{3,2}\mathcal{M}_{\theta_{2}}\widetilde{T}_{2,1}\mathcal{M}_{\theta_{1}}\widetilde{T}_{1,0}\right)[\rho_{\mathbb{S},0}].
\end{cases}
\end{equation*}
\end{small}

In addition, let us observe that also the multi-time statistics for $\mathbb{S}$, given by the joint probability distribution in Eq.(\ref{joint_prob}), can be expressed by means of these recursive relations. Indeed, by using again Eq.\,(\ref{TTM_measure_2}), one has that
\begin{equation}
q_k = {\rm Tr}\left[\mathcal{M}_{\theta_k}\underline{\widetilde{T}}_{k}\underline{\mathcal{M}}_{\theta_{k - 1}}\underline{\Upsilon}_{k - 1}[\rho_{\mathbb{S},0}]\right],
\end{equation}
$\forall k = 1,\ldots,m$, where the first terms of the recursion for $q_k$ are given by
\begin{equation*}
\begin{cases}
q_0 = {\rm Tr}\left[\rho_{\mathbb{S},0}\right] = 1 \\
q_1 = {\rm Tr}\left[\mathcal{M}_{\theta_1}\widetilde{T}_{1,0}[\rho_{\mathbb{S},0}]\right] \\
q_2 = {\rm Tr}\left[\mathcal{M}_{\theta_2}\widetilde{T}_{2,0}[\rho_{\mathbb{S},0}]\right] + {\rm Tr}\left[\mathcal{M}_{\theta_2}\widetilde{T}_{2,1}\mathcal{M}_{\theta_1}\widetilde{T}_{1,0}[\rho_{\mathbb{S},0}]\right] \\
q_3 = {\rm Tr}\left[\mathcal{M}_{\theta_3}\widetilde{T}_{3,0}[\rho_{\mathbb{S},0}]\right] + {\rm Tr}\left[\mathcal{M}_{\theta_3}\widetilde{T}_{3,1}\mathcal{M}_{\theta_1}\widetilde{T}_{1,0}[\rho_{\mathbb{S},0}]\right] \\
\,\,\,\,\,\,\,+\,\,{\rm Tr}\left[\mathcal{M}_{\theta_3}\widetilde{T}_{3,2}\mathcal{M}_{\theta_2}\widetilde{T}_{2,0}[\rho_{\mathbb{S},0}]\right] \\
\,\,\,\,\,\,\,+\,\,{\rm Tr}\left[\mathcal{M}_{\theta_3}\widetilde{T}_{3,2}\mathcal{M}_{\theta_2}\widetilde{T}_{2,1}\mathcal{M}_{\theta_1}\widetilde{T}_{1,0}[\rho_{\mathbb{S},0}]\right].\label{eq:qt}
\end{cases}
\end{equation*}

Finally, also the set $\{\widetilde{T}\}$ of stochastic TTs admits a hierarchic structure, very similar to that of Eq.\,(\ref{hierarchy}), based on a generalization of the CPTP maps $\Gamma$s. The main difference is that now we need a map acting on the reduced dynamics of $\mathbb{S}$ at each $t_k$ after that a quantum measurement has been performed on the system. Thus, such map will depend on the resulting post-measurement environmental state. In particular, following the same construction as in \cite{Modi_Reconstructing_2018}, we get
\begin{equation}\label{stoc-TT-hierarchy}
\widetilde{T}_{k,k-n} =\widetilde{\Gamma}_{k|k-n} - \sum_{j=1}^{n-1}\widetilde{T}_{k,k-j}\mathcal{M}_{\theta_{k-j}}
\widetilde{\Gamma}_{k-j|k-n},
\end{equation}
where the CPTP maps
\begin{equation}\label{extra2}
\widetilde{\Gamma}_{k|k-n}[\sigma_{\mathbb{S}}]={\rm Tr}_{\mathbb{E}}[\mathcal{U}_{t_{k-n}:t_{k}}[\sigma_{\mathbb{S}}\otimes\widetilde{\sigma}_{\mathbb{E},k-n}]]
\end{equation}
also include a dependence on the whole sequence of outcomes, fixing the environmental state $\widetilde{\sigma}_{\mathbb{E},k-n}$ (which motivates the use of the tilde).

\subsection{Conditional transfer tensors and memory effects}

The relation in Eq.\,(\ref{stoc-TT-hierarchy}) allows us to bring forward the connection between TTs and CP-divisibility to the stochastic level that describes the effects sequential measurements.

In particular, let us consider the situation where, for any sequence of outcomes, all the stochastic TTs $\widetilde{T}_{k,k-n}$ are equal to 0 for $n\geq 2$. The stochastic TTs hierarchy of Eq.\,(\ref{stoc-TT-hierarchy}) implies that each $1$-step transfer tensor $\widetilde{T}_{k,k-1}$ is equal to the corresponding $1$-step $\widetilde{\Gamma}$, i.e., $\widetilde{T}_{k,k-1}=\widetilde{\Gamma}_{k|k-1}$, which by definition is completely positive. In addition, recursively applying $\widetilde{\Phi}_k = \sum_{j = 0}^{k - 1}\widetilde{T}_{k,j}\mathcal{M}_{\theta_{j}}\widetilde{\Phi}_j$, we get
\begin{equation}
\widetilde{\Phi}_k = \widetilde{T}_{k|k-1}\mathcal{M}_{\theta_{k-1}}\widetilde{T}_{k-1|k-2}\mathcal{M}_{\theta_{k-2}}\cdots\widetilde{T}_{j+1|j}\mathcal{M}_{\theta_{j}}\widetilde{\Phi}_{j}
\end{equation}
for any $k\geq j\geq 0$. Therefore, we have
\begin{equation}\label{eq-theorem-2}
\widetilde{\Phi}_k = \widetilde{\Gamma}_{k|k-1}\mathcal{M}_{\theta_{k-1}}\widetilde{\Gamma}_{k-1|k-2}\mathcal{M}_{\theta_{k-2}}\cdots\widetilde{\Gamma}_{j+1|j}
\mathcal{M}_{\theta_{j}}\widetilde{\Phi}_{j},
\end{equation}
which implies that for any $k\geq j\geq 0$ there is a (conditional) CP map $\widetilde{\mathcal{E}}_{k,j}$ such that
\begin{equation*}
\widetilde{\Phi}_k= \widetilde{\mathcal{E}}_{k,j}\widetilde{\Phi}_j
\end{equation*}
i.e., the family of maps $\widetilde{\Phi}_k$ is CP-divisibile\,\cite{foot3}. We conclude that, if all the stochastic TTs $\widetilde{T}_{k,k-n}$ are equal to 0 for $n\geq 2$,
the corresponding conditional dynamics $\{\widetilde{\Phi}_{k}\}_{k=1,\ldots,m}$ is CP-divisible for any fixed sequence of outcomes.

As one can directly check from the definition in Eq.\,(\ref{eq:phit}), the condition in Eq.\,(\ref{eq-theorem-2}) is satisfied whenever the global state after the measurements is a product-state, e.g., if the set $\{\mathcal{M}_{\theta_{k}}\}$ describes projective measurements of a non-degenerate system's observables. This means that in these situations the conditional dynamics $\widetilde{\Phi}_k$ will be automatically CP-divisible, simply due to the kind of employed measurement. As we will see in the next paragraph by means of an explicit example, CP-divisibility of the conditional dynamics, and in particular the validity of Eq.\,(\ref{eq-theorem-2}), does not imply that only one-step conditional TTs are different from zero.

The stochastic TTs $\widetilde{T}_{k,k-n}$ with $n\geq 2$ enclose the influence of the possible correlations resulting both from the system-environment interaction and the measurement on the open system up to a time step $j$ on the conditional dynamics at the later time step $k$. In general, such correlations depend on the whole sequence of measurement outcomes performed up to the $j-$th time instant, so that their influence on the subsequent conditional dynamics can be read as a signature of memory in the multi-time statistics defined by Eq.\,(\ref{joint_prob}). However, even in the case where Eq.\,(\ref{eq-theorem-2}) holds, some memory can be present in the conditional dynamics, and hence in the multi-time statistics. In fact, the post-measurement environmental state $\widetilde{\sigma}_{\mathbb{E}}$, which defines the conditional maps $\widetilde{\Gamma}$ in Eq.\,(\ref{extra2}), might (and in general will) depend on the sequence of outcomes. As we will see in the next paragraph, the proper description of such memory effects can require non-zero TTs $\widetilde{T}_{k,k-n}$ with $n\geq 2$.

Finally, it is worth observing that, if in addition to the validity of Eq.\,(\ref{eq-theorem-2}) the dependence of $\widetilde{\sigma}_{\mathbb{E}}$ on the previous outcomes can be neglected (i.e., no memory of the sequence of measurements is left), Eq.\,(\ref{eq-theorem-2}) automatically implies the quantum regression theorem \cite{Lax1968,vanKampen1992,Carmichael1993,Gardiner2004,GuarnieriPRA2014}. Essentially, the latter means that the whole multi-time statistics can be reconstructed only from the initial open-system state and open-system dynamical maps that are not conditioned on the sequence of outcomes. Similar considerations have been recently discussed also in\,\cite{Budini2018}.

\subsection{Spin-boson case study}

In this paragraph, we show an application of the previously introduced stochastic transfer tensors, with the aim to illustrate the capability of the method in giving a quantitative account of the memory effects both in the conditional dynamics and in the multi-time statistics. In addition, the use of TTs also allows to compare on a similar footing the memory effects in the presence of different kinds of measurements, or even no intermediate measurements at all.

We consider a single spin in interaction with $5$ quantum harmonic oscillators and monitored by a sequence of $M=5$ quantum measurements at regular intervals $\Delta$ with $t_k = k\Delta$ and $k=1,\ldots,M$. The Hamiltonian $H$ of the composite system is given by
\begin{equation}
  H = \frac{1}{2}\sigma^{z}_{\mathbb{S}}+\sum_{k=1}^{5}\omega_{\mathbb{E},k}b_{\mathbb{E},k}^{\dagger}b_{\mathbb{E},k} + \sum_{k=1}^{5}g_{k}(\sigma_{\mathbb{S}}^{-}b_{\mathbb{E},k}^{\dagger} + \sigma_{\mathbb{S}}^{+}b_{\mathbb{E},k})
\end{equation}
where $\sigma^{z}$ is the Pauli matrix in the $z$ direction, and $b_{\mathbb{E}}^{\dagger}, b_{\mathbb{E}}$ and $\sigma_{\mathbb{S}}^{+}, \sigma_{\mathbb{S}}^{-}$ denote the raising and lowering operators associated respectively to each harmonic oscillator of $\mathbb{E}$ and the spin $\mathbb{S}$. Moreover, regarding the frequencies $\omega_{\mathbb{E},k}$ of the oscillators and the interaction couplings $g_{k}$, for the numerical simulations here performed we have chosen the values $\{\omega_{\mathbb{E},k}\}_{k=1}^{5} \approx (1.99,0.73,0.89,2.04,1.58)$ and $\{g_{k}\}_{k=1}^{5} \approx (1.67,1.32,2.15,2.70,1.07)$. All these values are expressed in units ensuring that $\hbar=1$. The former have been uniformly sampled from the interval $[0,5]$, while the latter are sampled from the probability distribution $\alpha g\exp(-g/\beta)$, with $\alpha=1$ and $\beta=2$. The parameters $\alpha$ and $\beta$ tune the intensity, respectively, of the interaction between $\mathbb{S}$ and $\mathbb{E}$ and of the value of $g$ corresponding to the peak of the distribution. Note that this model accounts for both decoherence and dissipation in the dynamics of $\mathbb{S}$ due to the interaction with the external environment, yet with the form of the interaction preserving the total number of excitations. On the other hand, the excitations are generally not preserved by the action of the measurements. In our simulations, each measurement at $t=t_k$ is provided by the POVM operators\,\cite{AddisPRA2016}
\begin{equation}\label{eq:povm}
F_{\pm} \equiv (1-\lambda)|\pm\rangle\!\langle \pm|+\frac{\lambda}{2}\mathbbm{1}_{\mathbb{S}}\,,
\end{equation}
satisfying the normalization condition $\sum_{\theta_k \in \{+,-\}}F_{\theta_k} = \mathbbm{1}_{\mathbb{S}}$ for any $k$, where $|\pm\rangle \equiv (|0\rangle\pm|1\rangle)/\sqrt{2}$ denote the eigenstates of the Pauli matrix $\sigma^{x}$. The measurement-induced state transformations are ruled by the corresponding L{\"u}ders instrument, with maps
\begin{equation}
\mathcal{M}_{\pm}[\rho_k] = {\rm Tr}_{\mathbb{E}}\left[(\sqrt{F_{\pm}} \otimes \mathbbm{1}_{\mathbb{E}})\rho_{k}(\sqrt{F_{\pm}} \otimes \mathbbm{1}_{\mathbb{E}})\right]
\end{equation}
so that the post-measurement spin state $\widehat{\rho}_{\mathbb{S},k}$ collapses in one of the quantum states $\widehat{\rho}_{\mathbb{S},k}^{(\pm)}\equiv \mathcal{M}_{\pm}[\rho_k]/{\rm Tr}_{\mathbb{S}}[\mathcal{M}_{\pm}[\rho_k]]$. Such a choice for the measured observable is motivated by the fact that, by changing the parameter $\lambda$, it is possible to analyze the case without measurements ($\lambda=1$), the one with projective measurements ($\lambda=0$) and all the intermediate cases ($0<\lambda<1$) corresponding to a partial collapse of the spin wave-function.

The time interval $\Delta$ between measurements are taken equal to $1,2,3$ (in natural units allowing for $\hbar=1$), and the evolution of the composite system is evaluated by starting from an initial product-state of the ground state of $\mathbb{S}$ and a pure state of $\mathbb{E}$ (not its ground). Specifically, the ground state of $\mathbb{S}$ is obtained by diagonalizing $H_{\mathbb{S}}=\sigma^{z}_{\mathbb{S}}/2$, while the initial reduced state of $\mathbb{E}$ by diagonalizing a random perturbation of $H_{\mathbb{E}}=\sum_{k}\omega_{\mathbb{E},k}b_{\mathbb{E},k}^{\dagger}b_{\mathbb{E},k}$ where the perturbation is taken to preserve the Hermitianity of the resulting operator. Lastly, to get the reduced states of $\mathbb{S}$ and $\mathbb{E}$, the corresponding lowest energy eigenvectors are selected. Conversely, the hierarchy of transfer tensors is derived by numerically implementing Eq.\,(\ref{stoc-TT-hierarchy}). The latter is based on the computation of the maps $\widetilde{\Gamma}_{k|k-n}$, which are all obtained by propagating an initial product-state composed by the reduced states of $\mathbb{S}$ and $\mathbb{E}$ after each possible quantum measurement at the time instants $t_k$.

\begin{figure}[t!]
\centering
\includegraphics[width=0.975\columnwidth]{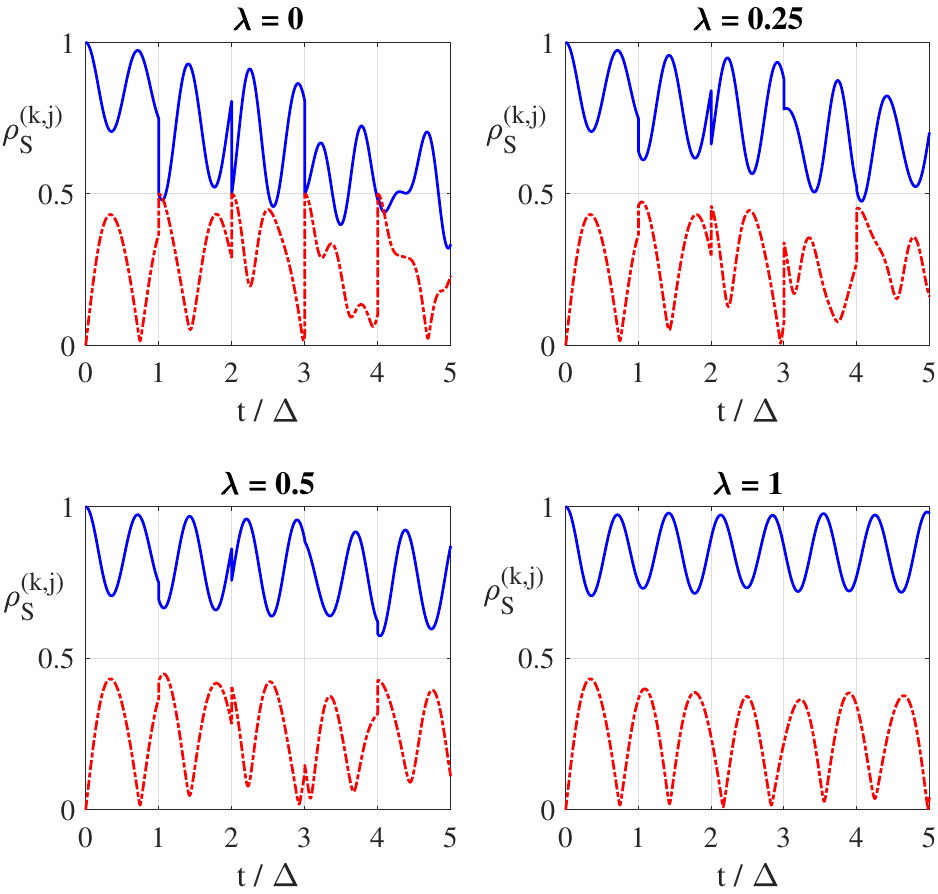}
\caption{Spin-boson dynamics: Ground state population $\rho_{\mathbb{S}}^{(1,1)}$ (blues solid lines) and coherence terms $|\rho_{\mathbb{S}}^{(1,2)}|$ (red dash-dotted lines) of the single spin $\mathbb{S}$ as a function of time and $4$ distinct $\lambda$ values for $\Delta = 1$ in natural units. Depending on the value of $\lambda$ that fixes the intermediate measurement operator, the coherence and population of the spin has a different behaviour. In particular, revivals occur for $\lambda=1$, while a more and more pronounced damping is observed for decreasing values of $\lambda$. Indeed, sudden changes in the population and coherence terms denote the action of the measurements, and they
are more evident in the case of projective measurements ($\lambda=0$).}
\label{fig:rho_terms_spin}
\end{figure}

\begin{figure*}[t!]
\centering
\includegraphics[scale=0.84]{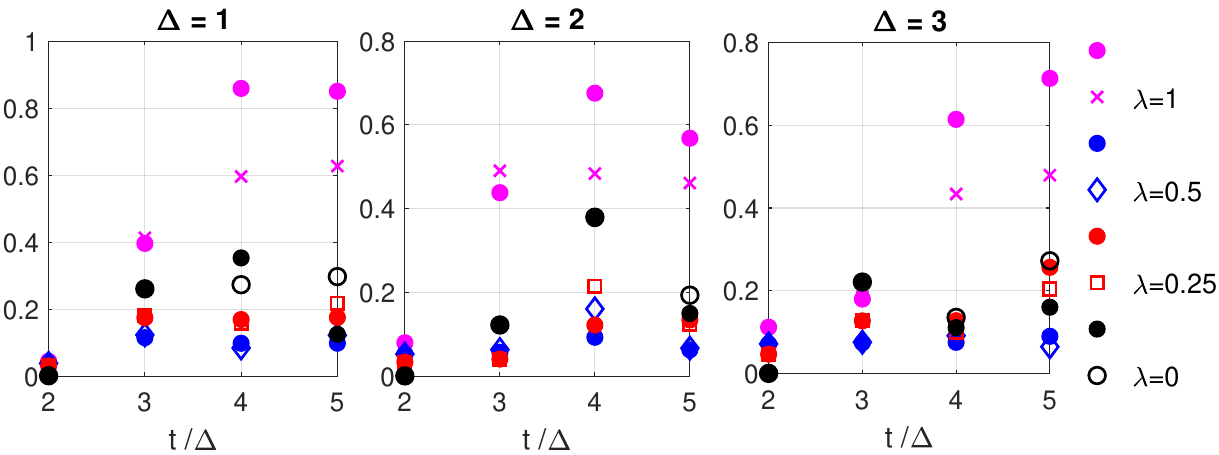}
\caption{Comparison between the $L_2$-norm of the stochastic TTs $\widetilde{T}_{k,0}$ (coloured dots) and the figure of merit $\mathcal{D}_k$ (circles, squares, diamonds and x-marks) for $k=2,\ldots,5$ (corresponding to the discrete time instants $t_2,\ldots,t_5$) and $\lambda=0,0.25,0.5,1$ (identified, respectively, by the colour black, red, blue and magenta). The former quantifier takes into account the influence of the $k$-step TTs all referring to $t=t_{0}$, while the latter sums up the impact of all the involved $k-\ell$ steps TTs with $\ell=0,\ldots,k-2$ and they both account for the non-Markovianity of the implemented quantum dynamics. If all the stochastic TTs $\widetilde{T}_{k,k-n}$ are equal to 0 for $n\geq 2$, then both quantifiers are equal to zero (the opposite does not generally hold).}
\label{fig:fig1}
\end{figure*}

In Fig.\,\ref{fig:rho_terms_spin} we plot both the population and coherence terms of the single spin $\mathbb{S}$ as a function of time and $\lambda$ values ($\lambda \in \{0,0.25,0.5,1\}$) for $\Delta = 1$ (all in natural units). As one can observe from the figure, the spin quantum coherence is maintained oscillating over time in case of no applied quantum measurements ($\lambda=1$), and such behaviour of repeated ``coherence revivals'' is originated by the interaction of $\mathbb{S}$ with the only environment $\mathbb{E}$ composed of 5 quantum harmonic oscillators. Conversely, in case also the external measurement apparatus plays a role (i.e., the open quantum system is also measured at consecutive time instants) the quantum coherence of the spin $\mathbb{S}$ may be increasing or decreasing depending on the value of $\lambda$ that corresponds to different intermediate measurement operators (at $\Delta$ fixed). Overall, concerning the effect of applying the single quantum operation, a quantum measurement tend to recreate coherence terms, unlike the interaction of the spin with the environment that on average is responsible to erase them. From Fig.\,\ref{fig:rho_terms_spin} one can clearly appreciate that monitoring the open quantum system $\mathbb{S}$ through a sequence of quantum measurements radically changes its reduced dynamics, as well as the corresponding non-Markovian behaviors. In particular, if the quantum coherence of $\mathbb{S}$ is damping over time, then also high-order step memory effects can be considered negligible. Otherwise, each state of $\mathbb{S}$ also depends on past contributions, even up to the initial instant in the extreme case.

Let us now analyze more quantitatively these considerations by introducing proper quantifiers of non-Markovianity. In this regard, in Fig.\,\ref{fig:fig1} we plot two quantifiers of the impact of the stochastic $n$-step TTs ($n\geq 2$) on the system's conditional dynamics. In each panel, we consider different POVMs, defined as in Eq.\,(\ref{eq:povm}) for $\lambda=0,0.25,0.5,1$ and identified respectively by the colours black, red, blue and magenta, as a function of the discrete time instants $t_k$ with $k=1,\ldots,5$. The three panels correspond to three different values of the time interval $\Delta$. The two quantifiers are the $L_2$-norm $\|\cdot\|_2$ of the stochastic transfer tensors $\widetilde{T}_{k,0}$ (coloured dots in the figure) and the figure of merit (circles, squares, diamonds and x-marks)
\begin{equation}
\mathcal{D}_k \equiv \frac{1}{k-2}\sum_{\ell=0}^{k-2}\|\widetilde{T}_{k,\ell}\mathcal{M}_{\theta_{\ell}}\widetilde{\Phi}_{\ell}\|_{2}\,.
\end{equation}
The choice of the $L_2$-norm is motivated by practical convenience (especially in view of the application to higher dimensional systems) and we leave for future studies the investigation about the use of different norms. Starting from the decomposition of the conditional map $\widetilde{\Phi}_{k}$ in correspondence of $t=t_k$ as given in Eq.\,(\ref{transfer_tensor_transformation_app}), one can observe that the former quantifier focuses on the role of the $k$-step TT for $k \geq 2$ (note, indeed, that
$\widetilde{\Phi}_{0} = \mbox{Id}$, i.e., no measurement is performed at the initial time instant), while the latter sums up the influence of all the involved $k-\ell$ steps TTs with $k \geq 2$ and $\ell=0,\ldots,k-2$ (thus, the $1$-step TT is not considered). If all the stochastic TTs $\widetilde{T}_{k,k-n}$ are equal to 0 for $n\geq 2$, then both quantifiers are equal to zero; the opposite does not generally hold.

Overall, from Fig.\,\ref{fig:fig1} we can observe that the quantities $\|\widetilde{T}_{k,0}\|_{2}$ and $\mathcal{D}_k$ are comparable; namely, in the considered parametric regime the $k$-th step TT provides with similar information on the memory effects influencing $\widetilde{\Phi}_{k}$, compared to the sum of all the contributions from the TTs with step larger than 1. They both indicate that non-Markovian behaviours, originated from the interaction of the spin both with the environment and the observer, become relevant for $t/\Delta \geq 3$ with all the considered values of $\Delta$. Moreover, both the non-Markovianity quantifiers reach their maximum values in the case where no intermediate measurement is performed (magenta dots and x-marks), corresponding in Fig.\,\ref{fig:rho_terms_spin} to a nearly periodic evolution of the coherence term $|\rho_{\mathbb{S}}^{(1,2)}|$ over time. Thus, comparing Figs.\,\ref{fig:rho_terms_spin} and \ref{fig:fig1}, we can deduce that such periodic behaviours are originated by high-order step memory effects, presumably even from $t_0$. While without applying any measurement both quantifiers typically increase with time, a more pronounced non-monotonic behavior is observed in the presence of intermediate measurements, which reduce the memory effects caused by the interaction between $\mathbb{S}$ and $\mathbb{E}$. Although a decreased impact of multi-step TTs in the presence of intermediate measurements (corresponding to a decreased impact of system-environment correlations on the conditional dynamics) might have been expected, the quantifiers do not necessarily reach their minimum values in the case of projective measurements ($\lambda=0$). Rather, the minimum values are reached more often for $\lambda=0.25$ or $\lambda=0.5$ (the behavior for these two POVMs is quite similar). However, as anticipated, $n$-step TTs for $n\geq 2$ can be non-zero also in the case of projective measurements, i.e., when Eq.\,(\ref{eq-theorem-2}) holds. Such TTs account for memory effects due to the dependence of the post-measurement environmental state on the previous outcomes. Even more, the latter can exceed the memory effects -- due to both system-environment correlations and changes in the environmental states -- occurring in the case of non-projective measurements, as quantified by $\|\widetilde{T}_{k,0}\|_{2}$ and $\mathcal{D}_k$.

\begin{figure}[t!]
\centering
\includegraphics[width=0.95\columnwidth]{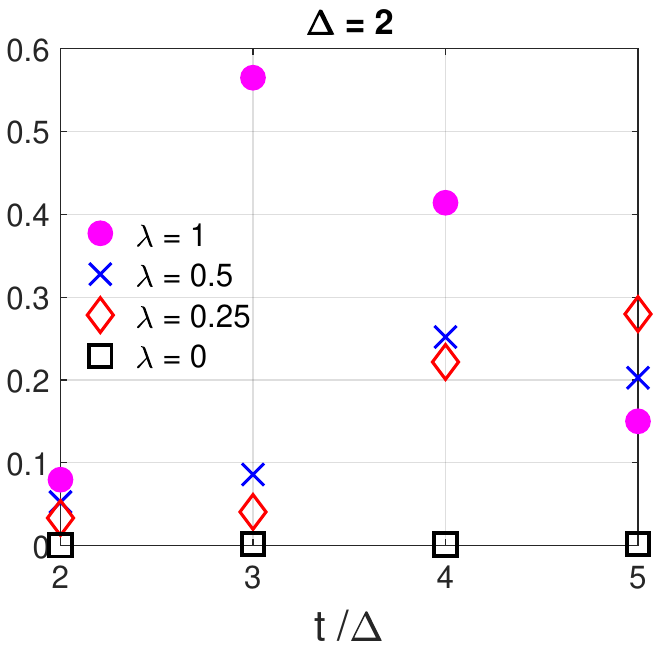}
\caption{$L_2$-norm of the difference between the left- and the right-hand-side of equation (\ref{eq-theorem-2}), namely $\|\widetilde{\Phi}_k - \widetilde{\Gamma}_{k|k-1}\mathcal{M}_{\theta_{k-1}}\widetilde{\Gamma}_{k-1|k-2}\mathcal{M}_{\theta_{k-2}}\cdots\widetilde{\Gamma}_{j+1|j}
\mathcal{M}_{\theta_{j}}\widetilde{\Phi}_{j}\|_2$, computed for $\Delta=2$ and $\lambda=0,0.25,0.5,1$ (magenta dots, blue x-marks, red diamonds, black squares, respectively) in correspondence of the discrete time-instants $t_k=k\Delta$ with $k\in\{2,3,4,5\}$. This represents the $L_2$-norm of the difference between the actual stochastic map $\widetilde{\Phi}_k$ and the CP-divisible map provided by the product-states of the dynamical evolution of the spin after each measurement. During the transient of the spin dynamics, Eq.\,(\ref{eq-theorem-2}) is violated as the value of $\lambda$ increases (thus, for less invasive intermediate measurements); however, such tendency can be reversed in case the spin approaches an equilibrium asymptotic state or a stable periodicity phase.}
\label{fig:fig2}
\end{figure}

In order to point out the memory effects originated by system-environment correlations, we consider the $L_2$-norm of the difference between the left- and the right-hand-side of Eq.\,(\ref{eq-theorem-2}), namely between the actual stochastic map $\widetilde{\Phi}_k$ (in correspondence of the $k$-th time instant $t_k$) and the CP-divisible map $\widetilde{\Gamma}_{k|k-1}\mathcal{M}_{\theta_{k-1}}\widetilde{\Gamma}_{k-1|k-2}\mathcal{M}_{\theta_{k-2}}\cdots\widetilde{\Gamma}_{j+1|j}\mathcal{M}_{\theta_{j}}\widetilde{\Phi}_{j}$ associated with the product-states after each measurement. The results are plotted in Fig.\,\ref{fig:fig2}. It is worth noting that Eq.\,(\ref{eq-theorem-2}) is satisfied only when projective measurements are performed, so that the composite system is actually in a product-state after each measurement and the conditional dynamics of $\mathbb{S}$ can be fully described by means of the composition of $1$-step maps $\widetilde{\Gamma}$. In this case, the computation of the operators $\widetilde{\Gamma}_{k|k-1}$ is enough by itself. However, as said, some $n$-step stochastic TTs with $n\geq2$ can be still different from zero, despite overall their combination is canceled to ensure the validity of Eq.\,(\ref{eq-theorem-2}).

As final remark, from Fig.\,\ref{fig:fig2} we can also observe that for $k=2,3,4$ the violation of Eq.\,(\ref{eq-theorem-2}) increases with the value of $\lambda$, i.e., when moving toward less invasive measurements; indeed, the largest violation is reached for the case without intermediate measurements. Moreover, quite interestingly, the situation is reversed in the correspondence of the last observed value of time, $t_k/\Delta=5$, for which $\lambda=0.25$ induces the largest violation, while the unconditional dynamics the smallest (apart, of course, from the case of projective measurements). The latter behaviour of the unconditional dynamics is usually originated by a tendency towards an equilibrium asymptotic state or to a stable periodicity phase. In our case study, being the environment composed by a finite number (five) of quantum harmonic oscillators, the simulated unconditional dynamics is practically periodic, with period approximately of $t=10$ (in natural units).

\section{Conclusion and outlook}\label{sec:ceo}

In this paper, we have studied the non-Markovian dynamics of open quantum systems and their multi-time statistics by means of the transfer-tensor approach,
which allowed us to treat the memory effects in the two different scenarios on a similar footing. After showing the connection between the hierarchy of the TTs and the divisibility of the dynamics, we have extended the definition of TTs to the case where the open system of interest undergoes quantum measurements at subsequent instants of time.
We have thus introduced a stochastic family of TTs, depending on the sequence of measurement outcomes, and a related hierarchy that captures how the multi-time statistics is influenced by both the system-environment correlations and the dependence of the current environmental state on the previous outcomes. Finally, we have defined two quantifiers of multi-time memory effects, which rely directly on the various contributions to the hierarchy of TTs, and we have investigated their behavior in a paradigmatic case study, comparing different kinds of intermediate measurements, as well as the case with no monitoring at all.

The precise relation between the memory effects present, respectively, in non-Markovian quantum dynamics and in their multi-time statistics remains still to be addressed. In particular, more quantitative and general results characterizing different kinds of multi-time statistics, such as those obeying the quantum-regression theorem, seem to be needed in order to distinguish the different sources of memory and to compare them with those leading to non-Markovian quantum dynamics.
Hopefully, our results will provide useful insights to look for such rigorous connections. More in general, our analysis suggests the usefulness of the TT approach also to deal with multi-time statistics; an example in this direction might be to use the hierarchy of stochastic TTs to evaluate the non-Markovianity along the so-called most probable trajectory of the system, namely the trajectory originated with higher probability by a sequence of quantum measurements\,\cite{WeberNature2014,GherardiniPRA2019}. In addition, the TT formalism addressed in this paper is also expected to provide a powerful tool to evaluate experimentally the degree of non-Markovianity \cite{LiEPL2019} in monitored quantum dynamics \cite{White2020,Xiang2021}, as well as to test the validity of the quantum regression theorem or possible generalizations \cite{Goan2011,Mccutcheon2016,deVegaRMP2017}.
One further advantage of the presented approach would be to track memory effects also in the reduced state evolution of the environment $\mathbb{E}$. This feature might be useful in those quantum technology devices (e.g., even the commercial ones now available online as IBM or Rigetti \cite{Morris2019,Perez2020,MartinaIBM2021}) in which the number of the environmental degrees of freedom is not much larger than the ones of the system $\mathbb{S}$ under analysis. Accordingly, in such a case, also the state of $\mathbb{E}$ is subjected to frequent and relevant changes due to the presence of $\mathbb{S}$, and the bi-directional exchange of information between the system and the environment likely leads to departures from Markovian dynamics and the quantum regression theorem.
Finally, it is worth also mentioning how stochastic transfer tensors may find application in modeling the interaction between a quantum system $\mathbb{S}$ and one or more external thermal baths, giving rise to thermalization processes in the large-time limit. During the transient of the dynamics, indeed, $\mathbb{S}$ may exhibit non-Markovian behaviours and non-trivial memory effects. In this regard, if for example we resort to quantum collision models \cite{RauPR1963,ScaraniPRL2002,ArisoyEntropy2019,CampbellEPL2021,CiccarelloArXiv2021}, a thermal bath is assumed of being composed by a large number of small subsystems, such that, at discrete time instants, the dynamics between the quantum system and the bath take place through successive ``collisions'' provided by pairwise short interactions. Similar thermalization effects are also obtained by monitoring $\mathbb{S}$ through a sequence of projective measurements, in the limit of many measurements \cite{GherardiniPRE2020}. Accordingly, in such a contexts, the TT formalism is expected to shed light in quantitative terms on the effective influence of transient non-Markovian dynamics, due to the interaction with thermal baths, within the quantum system under analysis in allowing for specific thermalization quantum processes.

\paragraph*{Note added:} During the completion of this work, the related paper \cite{Jorgensen2020} appeared, in which a generalization of the transfer-tensor formalism to multi-time measurements is introduced by means of the process-tensor description of the open-system evolution.

\begin{acknowledgements}
S.G. and F.C. were financially supported from PATHOS EU H2020 FET-OPEN grant No.\,828946, the Fondazione CR Firenze through the project Quantum-AI and UNIFI grant Q-CODYCES.
\end{acknowledgements}


\begin{thebibliography}{99}

\bibitem{WolfCMP2008}
M.M. Wolf, and J.I. Cirac. Dividing Quantum Channels. Commun. Math. Phys. {\bf 279}, 147-168 (2008).

\bibitem{BreuerPRL2009}
H.-P. Breuer, E.-M. Laine, and J. Piilo. Measure for the Degree of Non-Markovian Behavior of Quantum Processes in Open Systems.
Phys. Rev. Lett. {\bf 103}, 210401 (2009).

\bibitem{LainePRA2010}
E.-M. Laine, J. Piilo, and H.-P. Breuer. Measure for the non-Markovianity of quantum processes. Phys. Rev. A {\bf 81}, 062115 (2010).

\bibitem{Rivas2010PRL105}
A. Rivas, S.F. Huelga, and M.B. Plenio. Entanglement and non-Markovianity of quantum evolutions. Phys. Rev. Lett. {\bf 105}, 050403 (2010).

\bibitem{LiuNatPhys2011}
B.-H. Liu, L. Li, Y.-F. Huang, C.-F. Li, G.-C. Guo, E.-M. Laine, H.-P. Breuer, and J. Piilo. Experimental control of the transition from Markovian to non-Markovian dynamics of open quantum systems. Nat. Phys. {\bf 7}, 931-934 (2011).

\bibitem{ChruscinskiPRL2014}
D. Chruscinski, and S. Maniscalco. Degree of Non-Markovianity of Quantum Evolution. Phys. Rev. Lett. {\bf 112}, 120404 (2014).

\bibitem{Rivas_Review_2014}
A. Rivas, S.F. Huelga, and M.B. Plenio. Quantum non-{M}arkovianity: characterization, quantification and detection. Rep. Prog. Phys. {\bf 77}, 094001 (2014).

\bibitem{CarusoRMP2014}
F. Caruso, V. Giovannetti, C. Lupo, and S. Mancini. Quantum channels and memory effects. Rev. Mod. Phys. {\bf 86}, 1203 (2014).

\bibitem{Breuer_Review_2016}
H.-P. Breuer, E.-M. Laine, J. Piilo, and B. Vacchini. \textit{Colloquium}: Non-{M}arkovian dynamics in open quantum systems. Rev. Mod. Phys. {\bf 88}, 021002 (2016).

\bibitem{MullerSciRep2016}
M.M. M\"{u}ller, S. Gherardini, and F. Caruso. Stochastic quantum Zeno-based detection of noise correlations. Scientific Reports {\bf 6}, 38650 (2016).

\bibitem{deVegaRMP2017}
I. de Vega, and D. Alonso. Dynamics of non-Markovian open quantum systems. Rev. Mod. Phys. {\bf 89}, 015001 (2017).

\bibitem{Cialdi2017}
S. Cialdi,  M.A.C. Rossi, C. Benedetti, B. Vacchini, D. Tamascelli, S. Olivares, and  M.G.A. Paris. All-optical quantum simulator of qubit noisy channels. Appl. Phys. Lett. {\bf 110}, 081107 (2017).

\bibitem{PollockPRL2018}
F.A. Pollock, C. Rodríguez-Rosario, T. Frauenheim, M. Paternostro, and K. Modi. Operational Markov condition for quantum processes. Phys. Rev. Lett. {\bf 120}, 040405 (2018).

\bibitem{PollockPRA2018}
F.A. Pollock, C. Rodr{\'i}guez-Rosario, T. Frauenheim, M. Paternostro, and K. Modi, Non-Markovian quantum processes: Complete framework and efficient characterization. Phys. Rev. A {\bf 97}, 012127 (2018).

\bibitem{Budini2018}
A.A. Budini. Quantum Non-Markovian Processes Break Conditional Past-Future Independence. Phys. Rev. Lett. {\bf 121}, 240401 (2018).

\bibitem{CampbellPRA2018}
S. Campbell, F. Ciccarello, G.M. Palma, and B. Vacchini. System-environment correlations and Markovian embedding of quantum non-Markovian dynamics.
Phys. Rev. A {\bf 98}, 012142 (2018).

\bibitem{TarantoPRL2019}
P. Taranto, F.A. Pollock, S. Milz, M. Tomamichel, and K. Modi. Quantum Markov Order. Phys. Rev. Lett. {\bf 122}, 140401 (2019).

\bibitem{DoNJP2019}
H.-V. Do, C. Lovecchio, I. Mastroserio, N. Fabbri, F.S. Cataliotti, S. Gherardini, M.M. M\"{u}ller, N. Dalla Pozza, and F. Caruso. Experimental proof of Quantum Zeno-assisted Noise Sensing. New J. Phys. {\bf 21}, 113056 (2019).

\bibitem{CampbellNJP2019}
S. Campbell, M. Popovic, D. Tamascelli, and B. Vacchini. Precursors of non-Markovianity. New J. Phys. {\bf 21}, 053036 (2019).

\bibitem{Cialdi2019}
S. Cialdi, C. Benedetti, D. Tamascelli, S. Olivares, M. G. A. Paris, and B. Vacchini.
Experimental investigation of the effect of classical noise on quantum non-Markovian dynamics.
Phys. Rev. A {\bf 100}, 052104 (2019).

\bibitem{GherardiniPRR2020}
S. Gherardini, S. Marcantoni, and F. Caruso. Irreversibility mitigation in unital non-Markovian quantum evolutions.
Phys. Rev. Res. {\bf 2}, 033250 (2020).

\bibitem{MartinaArXiv2021}
S. Martina, S. Gherardini, and F. Caruso. Machine learning approach for quantum non-Markovian noise classification.
arXiv:2101.03221.

\bibitem{Li2018}
L. Li, M.J. W. Hall, H.M. Wiseman. Concepts of quantum non-Markovianity: a hierarchy. Phys. Rep. {\bf 759}, 1-51 (2018).

\bibitem{Breuer2002}
H. Breuer, and F. Petruccione. {\em The Theory of Open Quantum Systems} (Oxford University Press, Oxford, 2002).

\bibitem{Rivas2012}
{\'A}. Rivas, and S.F. Huelga. {\em Open Quantum Systems: An Introduction} (Springer, Berlin, 2012).

\bibitem{Vacchini2011}
B. Vacchini, A. Smirne, E.-M. Laine, J. Piilo, and H.-P. Breuer.
Markovian and non-Markovian dynamics in quantum and classical systems.
New J. Phys. {\bf 13}, 093004 (2011).

\bibitem{Smirne2013}
A. Smirne, A. Stabile, and B. Vacchini. Signatures of non-Markovianity in classical single-time probability distributions. Phys. Scr. {\bf T153}, 014057 (2013).

\bibitem{Milz2019}
S. Milz, M. S. Kim, F. A. Pollock, K. Modi. Completely Positive Divisibility Does Not Mean Markovianity.
Phys. Rev. Lett. {\bf 123}, 040401 (2019).

\bibitem{Smirne2017Coherence}
A. Smirne, D. Egloff, M.G. D{\'i}az, M.B. Plenio, and S.F. Huelga. Coherence and non-classicality of quantum Markov processes. Quantum Sci. Technol. {\bf 4}, 01LT01 (2019).

\bibitem{Strasberg2019}
P. Strasberg and M.G. D{\'i}az. Classical quantum stochastic processes. Phys. Rev. A. {\bf 100},
022120 (2019).

\bibitem{MilzQuantum20}
S. Milz, F. Sakuldee, F.A. Pollock, and K. Modi. Kolmogorov extension theorem for (quantum) causal modelling and general probabilistic theories.
Quantum {\bf 4}, 255 (2020).

\bibitem{SmirneArxiv2019}
A. Smirne, T. Nitsche, D. Egloff, S. Barkhofen, S. De, I. Dhand, C. Silberhorn, S.F. Huelga, and M.B. Plenio. Experimental Control of the Degree of Non-Classicality via Quantum Coherence. Quantum Sci. Technol. {\bf 5} 04LT01 (2020).

\bibitem{Diaz2020}
M.G. D{\'i}az, B. Desef, M. Rosati, D. Egloff, J. Calsamiglia, A. Smirne, M. Skotiniotis, and S. F. Huelga.
Accessible coherence in open quantum system dynamics. Quantum {\bf 4}, 249 (2020).

\bibitem{Milz2020}
S. Milz, D. Egloff, P. Taranto, T. Theurer, M.B. Plenio, A. Smirne, S.F. Huelga. When is a non-Markovian quantum process classical? Phys. Rev. X {\bf 10}, 041049 (2020).

\bibitem{proceeding_IQIS}
S. Gherardini, A. Smirne, M.M. M{\"u}ller, and F. Caruso. Advances in sequential measurements and control of open quantum systems. Proceedings {\bf 12 (1)}, 11 (2019).

\bibitem{Cerrillo2014PRL112}
J. Cerrillo, and J. Cao. Non-Markovian dynamical maps: {N}umerical processing of open quantum trajectories. Phys. Rev. Lett. {\bf 112}, 110401 (2014).

\bibitem{RosenbachNJP2016}
R. Rosenbach, J. Cerrillo, S.F. Huelga, J. Cao, and M.B. Plenio. Efficient simulation of non-Markovian system-environment interaction.
New J. Phys. {\bf 18}, 023035 (2016).

\bibitem{Modi_Reconstructing_2018}
F.A. Pollock, and K. Modi. Tomographically reconstructed master equations for any open quantum dynamics. Quantum {\bf 2}, 76 (2018).

\bibitem{ChenPRApp2020}
Y.-Q. Chen, K.-L. Ma, Y.-C. Zheng, J. Allcock, S. Zhang, and C.-Y. Hsieh. Non-Markovian Noise Characterization with the Transfer Tensor Method.
Phys. Rev. Applied {\bf 13}, 034045 (2020).

\bibitem{NielsenBook2000}
M.A. Nielsen, and I.L. Chuang. {\em Quantum Computation and Quantum Information} (Cambridge University Press, Cambridge, 2000).

\bibitem{MagesanPRL2011}
E. Magesan, J.M. Gambetta, and J. Emerson.
Scalable and Robust Randomized Benchmarking of Quantum Processes.
Phys. Rev. Lett. {\bf 106}, 180504 (2011).

\bibitem{foot1}
In fact, Eq.(\ref{eq:aux}) implies the property named No Information Backflow in \cite{Li2018}, which is sufficient but not necessary for CP-divisibility, so that the latter cannot imply $T_{k,k-n}=0$ for $n\geq 2$.

\bibitem{Heinosaari2012}
T. Heinosaari and M. Ziman. \emph{The Mathematical Language of Quantum Theory} (Cambridge University Press, Cambridge, 2012).

\bibitem{LuchnikovPRL2020}
I.A. Luchnikov, S.V. Vintskevich, D.A. Grigoriev, and S.N. Filippov.
Machine learning non-Markovian quantum dynamics.
Phys. Rev. Lett. {\bf 124}, 140502 (2020).

\bibitem{foot3}
We use here the term CP-divisible referred to a family of maps that are not necessarily trace-preserving. Hence, we also relax the requirement that the intermediate
maps $\widetilde{\mathcal{E}}_{k,j}$ are trace preserving.

\bibitem{Lax1968}
M. Lax. Quantum Noise. XI. Multitime Correspondence between Quantum and Classical Stochastic Processes. Phys. Rev. {\bf 172}, 350 (1968).

\bibitem{vanKampen1992}
N.G. van Kampen, \textit{Stochastic Processes in Physics and Chemistry} (North-Holland, Amsterdam,1992).

\bibitem{Carmichael1993}
H. Carmichael, \textit{An Open Systems Approach to Quantum Optics} (Springer-Verlag, Berlin, 1993).

\bibitem{Gardiner2004}
C. W. Gardiner and P. Zoller, \textit{Quantum Noise: A Handbook of Markovian and Non-Markovian Quantum Stochastic Methods with Applications to Quantum Optics} (Springer, Berlin, 2004).

\bibitem{GuarnieriPRA2014}
G. Guarnieri, A. Smirne, and B. Vacchini. Quantum regression theorem and non-Markovianity of quantum dynamics. Phys. Rev. A {\bf 90}, 022110 (2014).

\bibitem{AddisPRA2016}
C. Addis, T. Heinosaari, J. Kiukas, E.-M. Laine, and S. Maniscalco. Dynamics of incompatibility of quantum measurements in open systems.
Phys. Rev. A {\bf 93}, 022114 (2016).

\bibitem{WeberNature2014}
S.J. Weber, A. Chantasri, J. Dressel, A.N. Jordan, K.W. Murch, and I. Siddiqi. Mapping the optimal route between two quantum states.
Nature (London) {\bf 511}, 570 (2014).

\bibitem{GherardiniPRA2019}
S. Gherardini. Exact nonequilibrium quantum observable statistics: A large-deviation approach. Phys. Rev. A {\bf 99 (6)}, 062105 (2019).

\bibitem{LiEPL2019}
C.-F. Li, G.-C. Guo, J. Piilo.
Non-Markovian quantum dynamics: What is it good for?
EPL {\bf 128}, 30001 (2019).

\bibitem{White2020}
G.A.L. White, C.D. Hill,  F.A. Pollock, L.C.L. Hollenberg, and K. Modi.
Demonstration of non-Markovian process characterisation and control on a quantum processor.
Nat. Comm. {\bf 11}, 6301 (2020).

\bibitem{Xiang2021}
L. Xiang, Z. Zong, Z. Zhan, Y. Fei, C. Run, Y. Wu, W. Jin, C. Xiao, Z. Jia, P. Duan, J. Wu, Y. Yin, G. Guo.
Quantify the non-Markovian process with intermediate projections in a superconducting processor.
arXiv:2105.03333.

\bibitem{Goan2011}
H.-S. Goan, P.-W. Chen, and C.-C. Jian.
Non-Markovian finite-temperature two-time correlation functions of system operators: Beyond the quantum regression theorem.
J. Chem. Phys. {\bf 12}, 124112 (2011).

\bibitem{Mccutcheon2016}
D.P.S. McCutcheon.
Optical signatures of non-Markovian behavior in open quantum systems.
Phys. Rev. A {\bf 93}, 22119 (2016).

\bibitem{Morris2019}
J. Morris, F.A. Pollock, and K. Modi.
Non-Markovian memory in IBMQX4.
arXiv:1902.07980.

\bibitem{Perez2020}
G. Garc{\'\i}a-P{\'e}rez, M.A.C. Rossi, and S. Maniscalco.
IBM Q Experience as a versatile experimental testbed for simulating open quantum systems.
npj Quantum Inf. {\bf 6}, 1 (2020).

\bibitem{MartinaIBM2021}
S. Martina, L. Buffoni, S. Gherardini, and F. Caruso.
Learning the noise fingerprint of quantum devices.
arXiv:2109.11405.

\bibitem{RauPR1963}
J. Rau. Relaxation phenomena in spin and harmonic oscillator systems.
Phys. Rev. {\bf 129}, 1880 (1963).

\bibitem{ScaraniPRL2002}
V. Scarani, M. Ziman, P. Stelmachovic, N. Gisin, and V. Buzek.
Thermalizing Quantum Machines: Dissipation and Entanglement.
Phys. Rev. Lett. {\bf 88}, 097905 (2002).

\bibitem{ArisoyEntropy2019}
O. Arisoy, S. Campbell, and \"{O}.E. M\"ustecaplio\ifmmode \breve{g}\else \u{g}\fi{}lu.
Thermalization of finite many-body systems by a collision model.
Entropy {\bf 21}, 1182 (2019).

\bibitem{CampbellEPL2021}
S. Campbell, and B. Vacchini.
Collision models in open system dynamics: A versatile tool for deeper insights? 
EPL {\bf 133 (6)}, 60001 (2021).

\bibitem{CiccarelloArXiv2021}
F. Ciccarello, S. Lorenzo, V. Giovannetti, and G.M. Palma.
Quantum collision models: open system dynamics from repeated interactions.
arXiv:2106.11974.

\bibitem{GherardiniPRE2020}
S. Gherardini, G. Giachetti, S. Ruffo, and A. Trombettoni.
Thermalization processes induced by quantum monitoring in multilevel systems.
Phys. Rev. E {\bf 104 (3)}, 034114 (2021).

\bibitem{Jorgensen2020}
M. R. J{\o}rgensen and F. A. Pollock. A discrete memory-kernel for multi-time correlations in non-Markovian quantum processes. Phys. Rev. A {\bf 102}, 052206 (2020).

\end{thebibliography}
\end{document}